\documentclass{emulateapj}

\usepackage{graphicx,xspace}

\newcommand\chandra{{\it Chandra}\xspace}
\newcommand\xmm{{\it XMM-Newton}\xspace}

\begin{document}

\title{A New Calculation of Ne IX Line Diagnostics}

\author{Randall K. Smith, Guo-Xin Chen, Kate Kirby, Nancy S. Brickhouse}
\affil{Harvard-Smithsonian Center for Astrophysics \\
60 Garden St., Cambridge, MA 02138; rsmith@cfa.harvard.edu}

\begin{abstract}
We describe the effect that new atomic calculations, including
fully-relativistic R-matrix calculations of collisional excitation
rates and level-specific dielectronic and radiative recombination
rates, have on line ratios from the astrophysically significant ion
\ion{Ne}{9}. The new excitation rates systematically change some
predicted \ion{Ne}{9} line ratios by 25\% at temperatures at or below
the temperature of maximum emissivity ($4\times10^{6}$\,K), while the
new recombination rates lead to systematic changes at higher
temperatures. The new line ratios are shown to agree with observations
of Capella and $\sigma^2$\,CrB significantly better than older line
ratios, showing that 25-30\% accuracy in atomic rates is inadequate
for high-resolution X-ray observations from existing spectrometers.
\end{abstract}



\section{Introduction}

X-ray spectral models for collisionally ionized plasmas have improved
steadily over the past 30 years. As spectral lines dominate the
emission from an X-ray emitting collisional plasma over a wide range
of temperatures \citep{CT69}, a relatively large number of lines must
be included to create an accurate model. However, most early X-ray
observations had limited spectral resolution, $R \equiv E/\Delta E
\sim 1-20$, and approximate calculations of individual transitions
were generally adequate \citep{RS77,Mewe72,LM72}. As laboratory and
observational methods have improved, more complete calculations have
been produced \citep{APEC01,SPEX,Chianti1}, although questions remain
as to whether or not these are sufficiently accurate for astrophysical
applications.  With the launches of \chandra and \xmm, individual
emission lines from a wide range of elements are regularly resolved
and used in diagnostics, especially from the grating instruments on
these missions \citep{Canizares00, Brinkman01}. The strongest lines
are from He-like and H-like ions and the Fe L-shell ions. Since both
terrestrial and astrophysical plasmas are dominated over a broad
temperature range by emission from the complex Fe L-shell ions, a
collaboration of atomic physicists established the Iron Project
\citep{IP93} to perform high-accuracy R-matrix calculations of these
ions. However, no equivalent systematic calculation of line
emissivities for all He-like or H-like ions exists. We present here an
initial study of \ion{Ne}{9} comparing new high accuracy calculations
of collisional and radiative transition rates to a number of existing
results.

\begin{figure}
\begin{center}
\includegraphics[totalheight=2.5in]{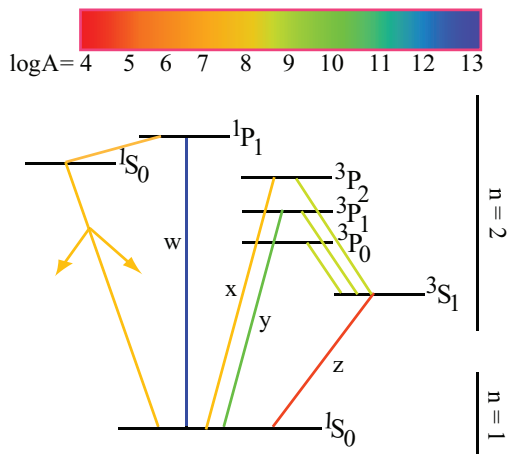}
\end{center}
\caption{Energy level diagram for the $n=1,2$\ levels of \ion{Ne}{9},
  with radiative transitions marked in colors that show the radiative
  transition rate (in s$^{-1}$). The strong $w, x, y,$\ and
  $z$\ transitions are marked \citep{Gabriel72}. The 2-photon
  transition is shown with two extra arrows.\label{fig:Elvl}}
\end{figure}

We began this project with He-like ions because they provide powerful
diagnostics on the physical conditions of the plasma. Neon is an
obvious choice as it is an astrophysically abundant element not
affected by dust depletion, with emission lines around 1 keV that can
be easily detected by both \chandra and \xmm. Furthermore,
experimental measurements of \ion{Ne}{9} spectra are available to test
the models \citep{Wargelin93}. In collisional equilibrium models, at
least 1\% of neon is He-like in plasmas with temperatures between
$5\times10^5$ and $10^7$\,K \citep{MM98}, and thus \ion{Ne}{9}
emission is present in numerous types of astrophysical sources.

The transitions from the n=2 He-like energy levels are
particularly favorable for use as diagnostics,
because their rates vary widely.  Figure~\ref{fig:Elvl} shows
the fine-structure levels arising from n=2 together
with the transitions: w, x, y, and z, with eight orders of
magnitude variation between the w (resonance line) and the
z (strongly forbidden line) transition rates.  The emission
lines from these transitions in He-like ions have been used in a
number of ways, including not only in the well-known G-ratio ($G
\equiv (x+y+z)/w$) for
temperature and ionization state and R-ratio ($R \equiv z/(x+y)$) for density \citep{GJ69},
described below, but also as diagnostics for photoionization
\citep{PD00}, photoexcitation \citep{Mewe1978, Kahn01}, and
non-Maxwellian velocity distributions \citep{Gabriel91}. Early
observations of He-like ion lines, including \ion{Ne}{9}, in solar
active region and flare X-ray spectra have provided general
diagnostics for density and temperature, but with complications due to
line blending, resonance scattering, and atomic data \citep{Doschek90,
  McKenzie87}. More recently, for stellar coronae, \ion{Ne}{9}
emission lines have been used to measure relative Ne abundances
\citep{Drake01, IIPeg01} and optical depths \citep{Ness03}, as well as
temperatures and densities \citep{Ayres01, Osten03,
  Ness03b}. Measurements of changing line ratios during stellar flares
are used to estimate the temperature and density in the heated plasma
\citep{Gudel02, Besselaar03, Raassen03}.

He-like emission line ratios have been calculated by a number of
authors, including \citet{BK00}, who considered both collisional and
photoionized plasmas, \citet{APEC01}, who 
focused on \ion{O}{7}, by \citet{Porquet01}, computing models for
\ion{C}{5}, \ion{N}{6}, \ion{O}{7}, \ion{Ne}{9}, \ion{Mg}{11}, and
\ion{Si}{13}. \citet{APEC01} showed that including higher-$n$\ states,
especially level-resolved dielectronic recombination to
high-$n$\ states, could significantly affect the G-ratio at higher
energies. \citet{Porquet01} included the effect of a radiation field
on the forbidden and intercombination lines, but these calculations
were based on distorted wave (DW) electron-impact excitation rate
calculations. \citet{ZS87} considered resonance effects manually
added to a DW calculation, but only for the $n=2$\ states. These were
included by \citet{APEC01}, with little resulting effect.

Despite advances in atomic calculation methods and the frequent use of
He-like diagnostics, \citet{BK00}, \citet{APEC01} and \citet{Porquet01}
all used rates taken from the DW calculations, primarily from
\citet{SGC83}. Although some modern R-matrix calculations are now
available \citep{B03, Delahaye06}, the compilation by \citet{SGC83} is
one of the few complete collections of He-like ion rates that includes
all ions of astrophysical interest at X-ray energies. A signal that
these data may not be adequate, however, came from a detailed study of
\ion{Ne}{9} lines from Capella, where \cite{Ness03b} found that the
G-ratio diagnostic suggested a temperature less than $2\times
10^6$\,K, despite the well-known and strong peak in Capella's emission
measure distribution at $6\times10^6$K. A similar discrepancy was
found by \citet{Testa04} from the G-ratios of \ion{O}{7} and
\ion{Mg}{11}.  Meanwhile, \citet{Osten03} inferred a low electron
density for $\sigma^2$\,CrB from the \ion{Ne}{9} R-ratio diagnostic,
despite finding larger values from other diagnostics.  These problems
suggested to us that resonances, which tend to increase the
collisional excitation rate of non-dipole transitions at low energies,
could be significant.  We knew these were not included in the existing
line ratio calculations of \citet{APEC01} that used the DW method.
However, the R-matrix method will automatically generate all relevant
resonances, and can be extended by carrying out a DW calculation for
larger principal quantum $n$\ as well.  A number of such calculations
exist, including \citet{B03}, \citet{Delahaye06}, and \citet{Chen06}.
For this work, we used the R-matrix calculation of Chen et al. (2006)
for \ion{Ne}{9}, which include the largest number of states (up to
$n=5$).

We briefly describe the astrophysical model in \S\ref{sec:method}, including
the R-matrix calculation and our sources for other data used in the
spectral calculation. We compare our results to previous work in
\S\ref{sec:results}, and discuss the effects on common spectroscopic
diagnostics in \S\ref{sec:discussion}.

\section{Method}\label{sec:method}

Once all the necessary atomic data have been compiled, calculating line
strengths for collisional plasmas is straightforward with the APEC
code \citep{APEC01}. APEC uses an explicit rate matrix formulation
including collisional and radiative transitions between levels, as
well as dielectronic and radiative recombination from the next higher
ionization state to the ground and excited levels of the ion under
consideration. The APEC code does not contain atomic data within the
code itself, but instead reads from a database of FITS files that can
be modified as needed. This makes comparisons between different
atomic calculations easy, as any changes in the database are entirely
independent of the code.

\subsection{Collisional Excitation Rate Calculation}

For $n\le5$, we use the collisional rate coefficients
of \citet{Chen06}. These electron collisional excitation
calculations use a fully-relativistic close-coupling approach. As
discussed in \citet{Chen06}, the additional resonance excitation
included in this calculation brings the model G-ratio values into
much better agreement with the laboratory 
results measured at the LLNL Electron-Beam Ion Trap (EBIT) experiment
\citep{Wargelin93}. 

\citet{APEC01} showed that high-$n$\ transitions (up to $n=10$)
can be significant to the high-temperature values of the G-ratio.
The variation appears to be largely due to dielectronic recombination
(see \S\ref{subsec:DR}) into high-$n$\ states, but for completeness
we added collisional excitation rates for levels $6 \le n \le 10$
calculated with the DW FAC code \citep{FAC}.

\subsection{Radiative recombination}

APEC initially used photoionization rates to calculate the rate of
radiative recombination to ground and excited states of \ion{Ne}{9},
via the principle of detailed balance, primarily from \citet{VY95} for
ground-state photoionization and \citet{CCB86} for excited-state
results. Recently \citet{Badnell06RR} calculated level-separated
radiative recombination rates for all atoms with $Z \le 30$\ and ion
stages up to and including Na-like ions recombining to Mg-like
ions. These data are significantly improved from the original rates in
APED, and they extend to higher individual $n$\ levels ($n \le
8$\ compared to $n \le 5$). We therefore use the \citet{Badnell06RR}
calculation in this paper. We note that a comparison of the existing
APEC results and \citet{Badnell06RR} shows that the two calculations
agree to better than 30\% for plasmas with temperatures between $10^4$
and $10^8$\,K for the rates to the individual $n=2$\ levels and the
total recombination rate to \ion{Ne}{9}.

\subsection{Dielectronic Recombination (DR)\label{subsec:DR}}

Dielectronic satellite recombination involves two electron
transitions. The incoming free electron, with energy $E_i$, recombines
into an excited level with energy $E_j$\ of ion $I$\ while also
exciting a bound electron into energy level $E_k$, leaving the ion in
a doubly-excited state. Unlike radiative recombination, this is a
resonant process since $E_i = E_j + E_k$. The doubly-excited ion then
relaxes when one electron transitions to a lower state (often a state
with the lowest unoccupied orbital) and either ejects the second
electron or emits a ``satellite line'' photon. This latter case is the
DR process, and the photon emitted appears as a red-shifted
``satellite'' of the transition that would occur without the presence
of the second excited electron. The ion is then left in an excited state,
which usually decays via a radiative transition.

All DR calculations include the explicit generation of many different
satellite transitions; depending upon the configurations included, the
number of transitions can range into the millions or more. Each
transition has a temperature-dependent rate that generates a
particular DR satellite line and leaves the ion in a specific excited
state. In practice the DR rates are summed either to generate the
total DR excitation rate for each excited state (eliminating
information about the satellite lines), or to obtain the total DR rate
for the ion (eliminating information about DR excitation). Of course
correctly calculating the output spectrum requires individual
transition data, which are difficult to obtain. At sufficiently high
resolution, the satellite lines can be separated from the primary
transitions, so for this project we ignore the satellite lines and use
DR data that have been summed to create level-specific recombination
rates.  However, as noted by \citet{BK00}, the G-ratio measured with
spectrometers with low- or medium-resolution may be affected by
blending of the primary lines with DR satellites.  For the DR data, we
use rates from the recent calculation of
\citet{Badnell06DR}\footnote{These can be downloaded from
http://www-cfadc.phy.ornl.gov/data\_and\_codes/aurost/aurost\_recomb/home.html}
, which include transitions up to $n=8$\ for recombinations to
\ion{Ne}{9}.

\section{Results}\label{sec:results}

We are now able to compare line ratio results from a number of
different calculations. We used four different calculations for this
paper. The {\bf Full}\ model uses the complete data set, as described
in \S\ref{sec:method}, while the {\bf Full (No Rec)}\ model uses the
complete data set excluding radiative and dielectronic recombination.
The {\bf ATOMDB v1.3.1}\ model uses the original ATOMDB v1.3.1
database \citep{APEC01}.  This model uses electron excitation rates
from \citet{SGC83}, radiative recombination rates derived from the
partial photoionization rates of \citet{CCB86}, dielectronic
recombination rates from \citet{VS78} and \citet{Saf00}, and radiative
transition rates from a number of sources \citep{Wiese96, DJ97, Lin77,
Fernley87, Drake88, Kelly87, SGC83}.  Details of which data source is used
for any particular transition is available on the ATOMDB website ({\tt
http://cxc.harvard.edu/atomdb/WebGUIDE/}.  The {\bf FAC} model
uses atomic data taken exclusively from the Flexible Atomic Code
\citep{FAC} and does not include recombination. These four cases allow
discrimination amongst changes due to the R-matrix calculation alone,
those due to a larger transition matrix, and those due to increased
consideration of recombination effects. We postpone to a later paper
consideration of effects due to innershell ionization from Ne$^{+7}$\
due to the small range of temperatures with both significant
Ne$^{+7}$\ ions and detectable \ion{Ne}{9} lines.  In all cases we
assume collisional ionization equilibrium, using the tables of \citet{MM98}.

\begin{figure}
\includegraphics[totalheight=2.5in]{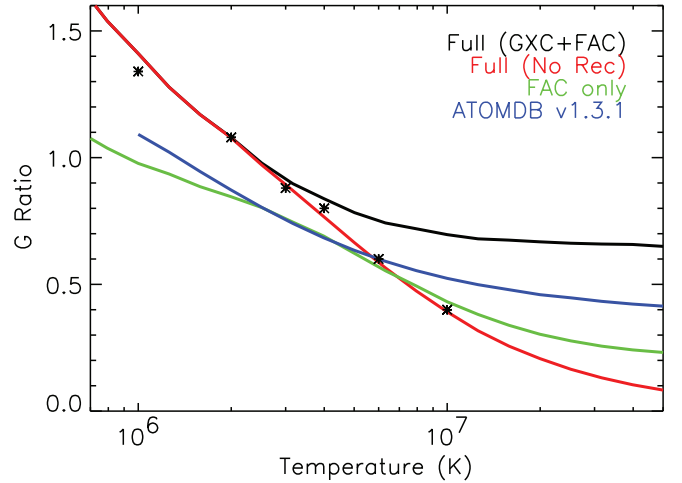}
\caption{\ion{Ne}{9}\ G-ratio as a function of temperature for four
  different calculations, at a density of 1 cm$^{-3}$. The starred
  points show data from EBIT measurements folded through a Maxwellian
  distribution.  The EBIT contained no Ne$^+{9}$\ during the
  experiment, and so the EBIT points (including an estimated 8\%
  error) should be compared to the red curve as no recombination
  occurred \citep{Wargelin93}; see \citet{Chen06} for more
  details. \label{fig:Gratio}}
\end{figure}

We focus on two line ratios commonly used for He-like ions, the
already-described G-ratio, which is a temperature and ionization state
diagnostic, and the R-ratio, a density diagnostic. In Figure~\ref{fig:Gratio} we show the G-ratio as a
function of temperature for our four different models. A number of
effects are immediately apparent: (1) The R-matrix results create a
systematically larger G-ratio, (2) the FAC results (which do not
include resonances, but do extend to $n\le10$) show a more complex
behavior which is generally closer to the current ATOMDB results, and
(3) including recombination systematically changes the result for
temperatures above $\sim 3\times10^6$\,K. The relative difference
between the {\bf Full}\ results and those in ATOMDB v1.3.1 is an
almost constant 25\%.

\begin{figure*}
\includegraphics[totalheight=2.5in]{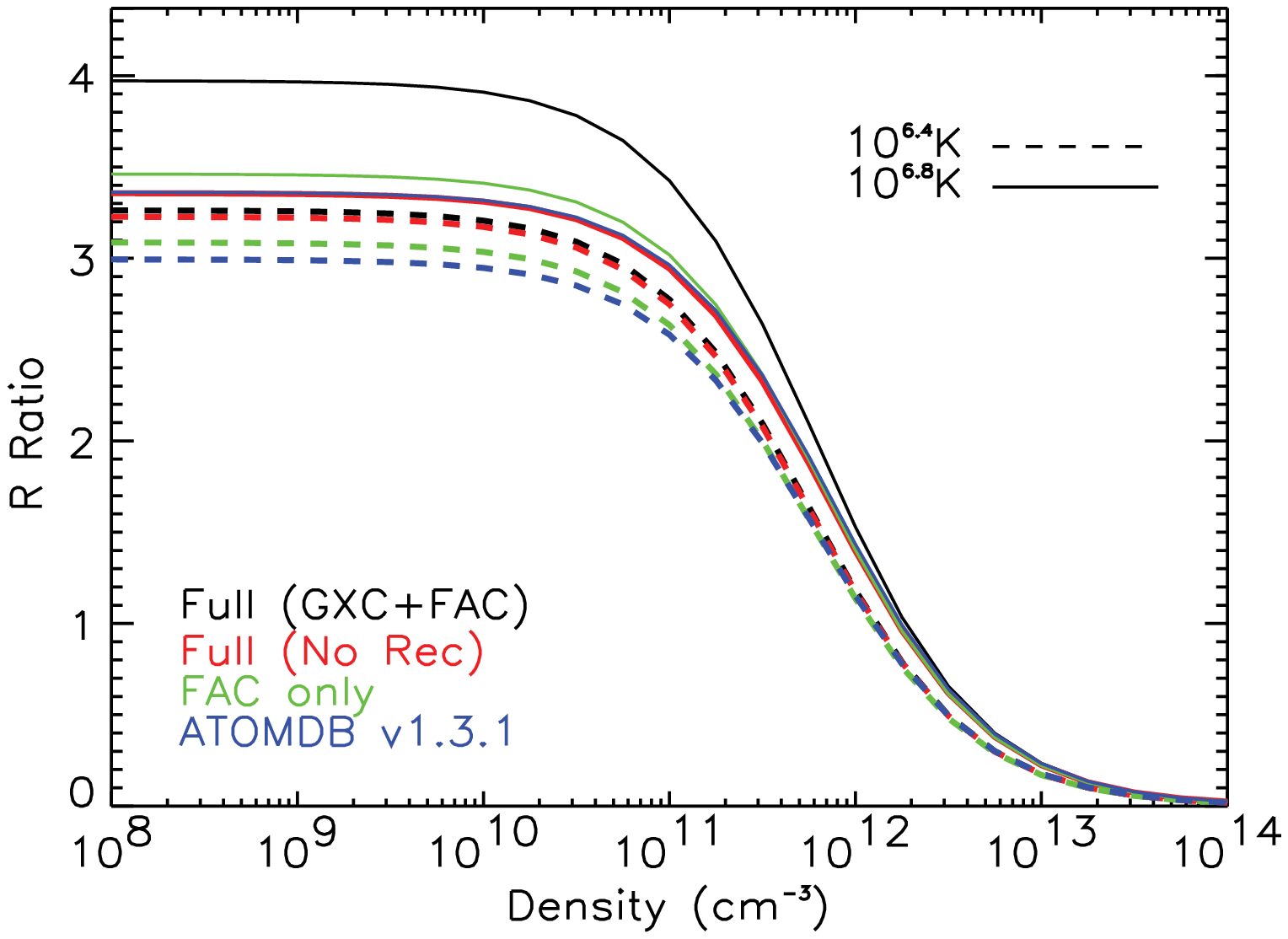}
\includegraphics[totalheight=2.5in]{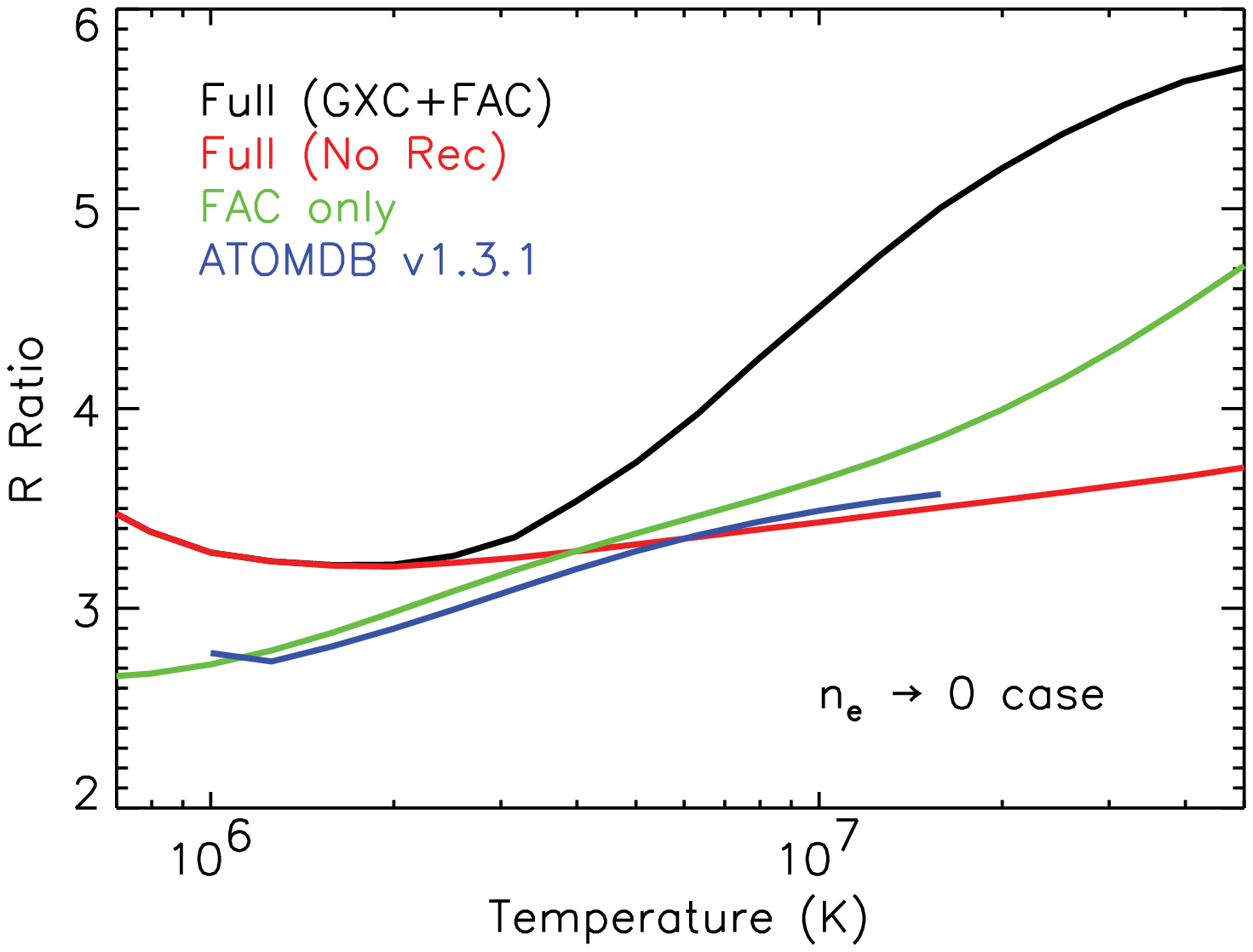}
\caption{[Left] \ion{Ne}{9}\ R-ratio as a function of density at
two different temperatures. [Right] \ion{Ne}{9}\ R-ratio as
a function of temperature at a density of 1 cm$^{-3}$.\label{fig:Rratio}}
\end{figure*}

Figure~\ref{fig:Rratio} shows the R-ratio as a function of electron
density at two different temperatures and as a function of
temperature (at an electron density of 1 cm$^{-3}$). As the density
increases, the slow radiative decay of the $1s2s{}^1$S$_{0}$\ to the
ground state (the $z$\ line) is eventually matched by electron
excitation to the $1s2p{}^3$P$_{1,2}$\ levels, which decay to the
ground state (creating the $x$ and $y$\ lines). For a given
temperature, the models are all quite similar at high densities, but
diverge from each other at lower densities, reflecting in part the
additional resonance contribution in the {\bf Full} and {\bf Full (No
Rec)} models. Also at the lower densities, the {\bf Full} and {\bf
Full (No Rec)} are similar for the lower temperature case (where very
little H-like neon is present) but differ significantly at the higher
temperature due to recombination. Above the low-density limit, as the
R-ratio begins to decline rapidly, the different models lead to
a factor of two difference in the inferred density.
Figure~\ref{fig:Rratio}[Right] shows that recombination produces a
significant increase in the R-ratio as the temperature increases.
However, at lower temperatures the biggest differences among the
models are due to differences between the R-matrix and DW calculations.

\section{Discussion and Conclusions}\label{sec:discussion}

The original goal of the project was to determine if the unexpected
results found by observers such as \citet{Osten03} and \citet{Ness03b}
were due to astrophysics or simply due to inadequate atomic models.
In both cases, the authors noted the issue but were unable to
distinguish between the two possibilities, and unfortunately, our
results confirm that there is no simple or quick method to select
between them. Our results show a difference between the DW and
R-matrix of $\sim 25\%$, within the oft-quoted error of
``approximately 30\%'' for atomic calculations. They also demonstrate
that this 30\% inaccuracy is inadequate for use with bright lines
observed by current X-ray satellites.

\begin{figure*}
\includegraphics[totalheight=2.5in]{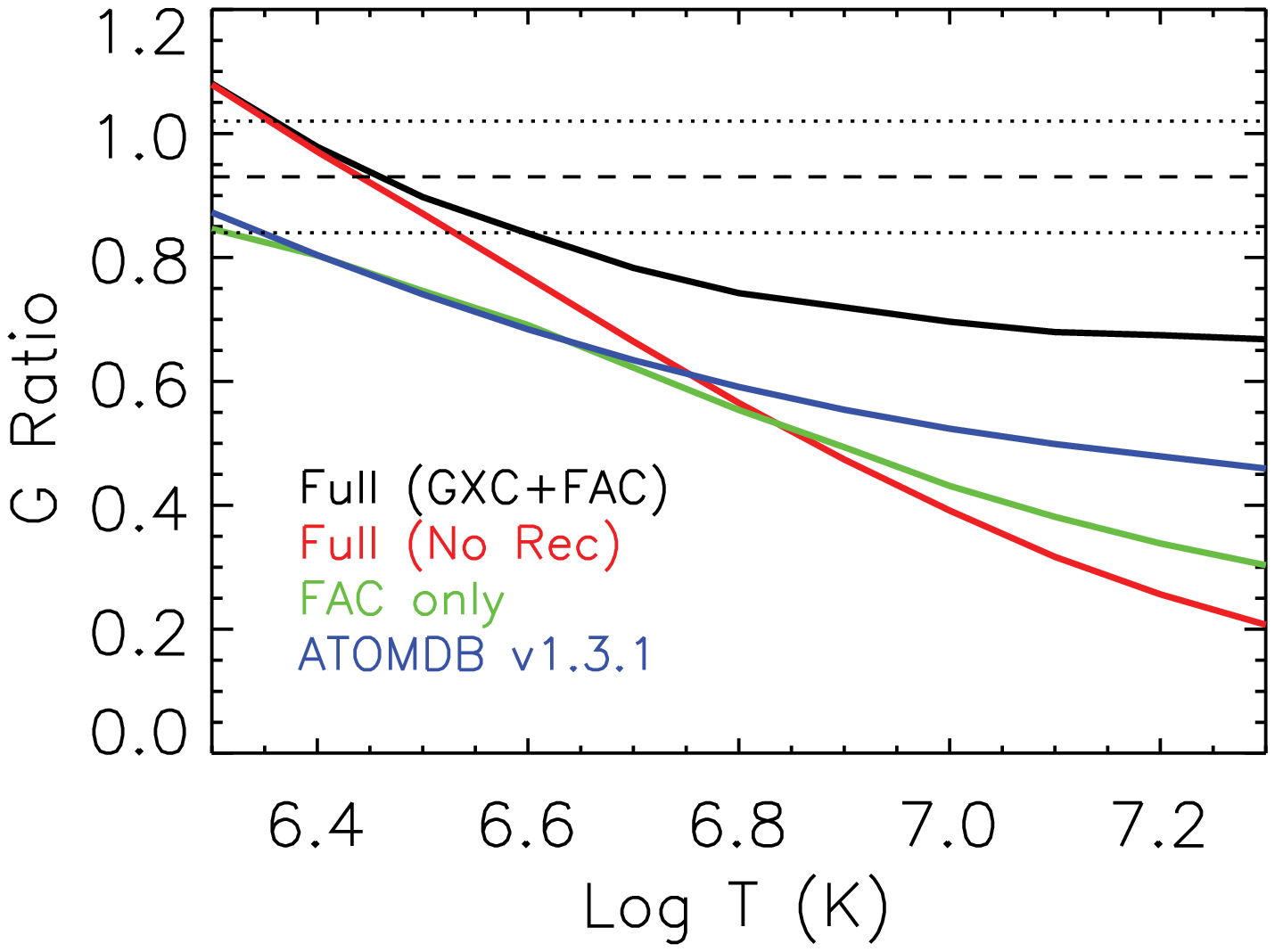}
\includegraphics[totalheight=2.5in]{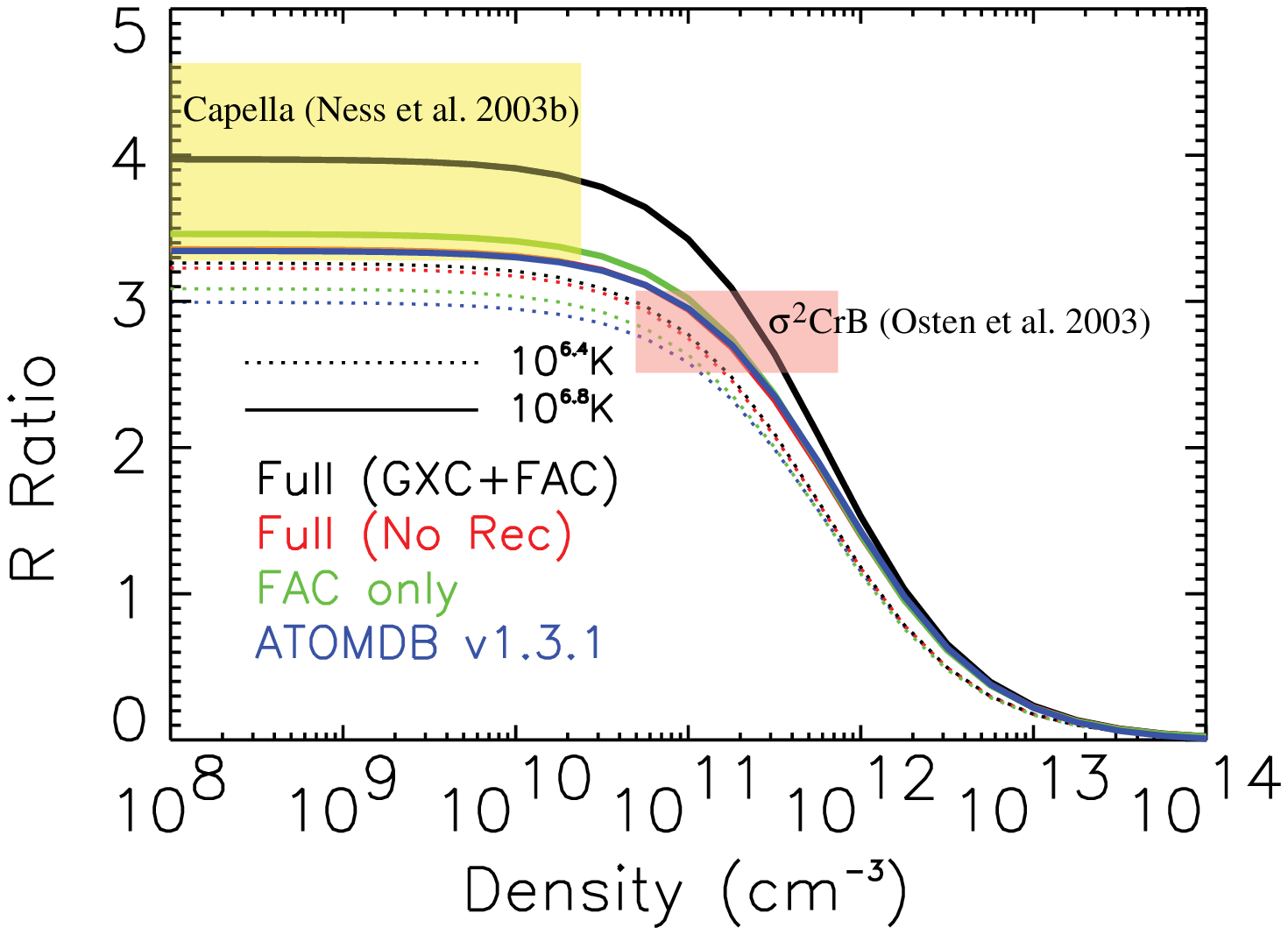}
\caption{[Left] \ion{Ne}{9}\ G-ratio with horizontal lines showing the
  measurement of G from Capella. The best-fit temperature is $\log T =
  6.46\pm0.14$, in much better agreement with the temperature of
  maximum emissivity for \ion{Ne}{9} ($\log T = 6.6$).  [Right]
  \ion{Ne}{9}\ R-ratio plotted for all four calculations, at
  $T=10^{6.4}$K\ and $T=10^{6.8}$K. Shown in yellow is the range of R
  measured from Capella \citep{Ness03b}, while the pink box shows the
  measurement from $\sigma^2$\,CrB by \citet{Osten03}. \label{fig:ratio_cmp}}
\end{figure*}

In Figure~\ref{fig:ratio_cmp}[Left] we show the \ion{Ne}{9} G-ratio in
the same format as Figure 11 of \citet{Ness03b}. The horizontal
dashed line shows the measured value from Capella (with errors shown
as dotted lines). As noted by \citet{Ness03b}, the result from ATOMDB
v1.3.1 barely agrees with the observed ratio, and only at a very low
temperature. Our new calculation predicts a temperature in the range
$\log T = 6.3 - 6.6$. Differential Emission Measure (DEM) fits of
Capella using X-ray and EUV spectra have consistently shown a strong
peak at $\log T \approx 6.8$\ \citep{Brickhouse95}, although with some
dispersion and emission at other temperatures. The agreement between
our results and laboratory measurements \citep{Chen06} suggests that
the remaining difference between the G-ratio and DEM temperatures may
well be due to the plasma conditions, and not to errors in atomic
physics.

Although the G-ratio is useful as a temperature diagnostic,
temperature can also be measured from the bremsstrahlung continuum or
from broad spectral fits. However, density diagnostics require
specific line ratios, and the strong lines emitted by He-like ions
make the R-ratio both useful {\it and}\
usable. Figure~\ref{fig:ratio_cmp}[Right] shows the ratio for all four
calculations, done at two temperatures, $\log T=6.4$\ and 6.8.  These
were chosen since they cover the range from the best-fit temperature
using the G-ratio and from DEM fits for both Capella \citep{Ness03b},
and, coincidentally, $\sigma^2$\,CrB \citep{Osten03}.  \citet{Ness03b}
noted that their value for the R-ratio in Capella ($3.97\pm0.67$,
shown as the yellow box) is only marginally in agreement with the
ATOMDB v1.3.1 values. We see here that the new calculation is now
consistent with the low-density limit.  Meanwhile, the lower value
found by \citet{Osten03} of $R=2.79\pm0.28$\ (shown as the pink box)
led to an inferred density at or near the low-density limit (less than
$5\times10^{10}$\,cm$^{-3}$), while the new results show a measurable
density from \ion{Ne}{9} in $\sigma^2$\,CrB, ranging from
$5\times10^{10} - 3.5\times10^{11}$\,cm$^{-3}$.

Our results show that an update for the He-like isosequence for all
astrophysically important elements is urgently needed, since the
existing data systematically underestimate key diagnostic line ratios.
This process is underway, with new calculations being done by our and
other groups. All available updates will be included in the next
revision (v2.0.0) of the ATOMDB.

\acknowledgements{We acknowledge support from the Chandra GO Theory
  program. Support for NSB was provided by NASA contract NAS8-03060 to
  the Smithsonian Astrophysical Observatory for the CXC.}

\end{document}